\documentstyle[aps,twocolumn,prb]{revtex}
\begin{document}

\input BoxedEPS
\SetTexturesEPSFSpecial
\HideDisplacementBoxes


\title{Relation between crystal and magnetic structures of the layered
manganites La$_{2-2x}$Sr$_{1+2x}$Mn$_2$O$_7$ ($0.30 \leq x \leq 0.50$)}

\author{M. Kubota}
\address{Neutron Scattering Laboratory, I.S.S.P., University of Tokyo, Tokai, Ibaraki,
319-1106, Japan}
\author{H. Fujioka and K. Hirota}
\address{CREST, Department of Physics, Tohoku University, Aoba-ku, Sendai, 980-8578,
Japan} 
\author{ K. Ohoyama} 
\address{Institute for Materials Research, Tohoku University, Aoba-ku, Sendai 980-8577,
Japan}
\author{ Y. Moritomo}
\address{Center for Integrated Research in Science and Engineering, Nagoya University,
Nagoya, 464-8601, Japan}
\author{H. Yoshizawa}
\address{Neutron Scattering Laboratory, I.S.S.P., University of Tokyo,  Tokai, Ibaraki,
319-1106, Japan}
\author{ Y. Endoh}
\address{CREST, Department of Physics, Tohoku University, Aoba-ku, Sendai, 980-8578,
Japan}

\date{February 21, 1999}

\twocolumn[\hsize\textwidth\columnwidth\hsize\csname @twocolumnfalse\endcsname
\maketitle

\begin{abstract}

Comprehensive neutron-powder diffraction and Rietveld analyses were carried  out to
clarify the relation  between the crystal and magnetic structures of
La$_{2-2x}$Sr$_{1+2x}$Mn$_2$O$_7$ ($0.30 \leq x \leq 0.50$).  The Jahn-Teller (JT)
distortion of Mn-O$_6$ octahedra, i.e., the ratio of the averaged apical Mn-O bond
length to the equatorial Mn-O bond length, is $\Delta_{JT}=1.042(5)$ at
$x=0.30$, where the magnetic easy-axis at low temperature is parallel to the
$c$ axis.  As the JT distortion becomes suppressed with increasing $x$, a  planar
ferromagnetic structure appears at $x \ge 0.32$, which is followed by a canted 
antiferromagnetic (AFM) structure at $x \ge 0.39$.  The canting angle
between neighboring planes continuously increases from 0$^{\circ}$ (planar ferromagnet:
$0.32 \leq x < 0.39$) to 180$^{\circ}$ (A-type AFM: $x=0.48$ where
$\Delta_{JT}=1.013(5)$).  Dominance of the A-type AF structure with decrease of JT
distortion can be ascribed to the change in the
$e_{g}$ orbital state from
$d_{3z^2-r^2}$ to $d_{x^2-y^2}$. 

\end{abstract}

]


It is now the received wisdom that the colossal magnetoresistance (CMR) discovered in
Perovskite Mn oxides R$_{1-x}$A$_{x}$MnO$_{3}$ (R: Rare-earth  ion, A: Alkaline-earth
ion) is originated from a complicated interplay of spin-charge-lattice degrees of
freedom.\cite{Urushi95,Millis95}  We can  control the hole concentration $x$ and the
effective band-width through the  tolerance factor.  This controllability makes this
system an ideal platform to study various phenomena related to CMR.  Recently, the
dimensionality of the  crystal structure has been focused upon.  In general, layered Mn
oxides are written as the Ruddelesden-Popper type system such as
(La,Sr)$_{n+1}$Mn$_n$O$_{3n+1}$.  By changing the number of MnO$_{2}$  sheets,
$n$, blocked with (La,Sr)$_{2}$O$_{2}$ layers, the effective dimensionality  can be
adjusted.  As for La$_{1-x}$Sr$_{x}$MnO$_{3}$ ($n=\infty$) and single-layered
La$_{1-x}$Sr$_{1+x}$MnO$_{4}$ ($n=1$), the magnetic phase diagrams in a wide $x$ range
were reported.\cite{Kaw96,Moritomo95}

Recently, Moritomo {\it et al.}\cite{Moritomo96} have synthesized a single crystal of
La$_{1.2}$Sr$_{1.8}$Mn$_2$O$_7$ ($n=2$) and found an extremely large
magnetoresistance.  This discovery has clarified the importance of effective
dimensionality in CMR, and was followed by intensive studies of
La$_{2-2x}$Sr$_{1+2x}$Mn$_2$O$_7$,\cite{Perring97,Kimura97,Argyriou97,Perring98}
abbreviated to LSMO327 in the present Communication.   Hirota {\it et
al.}\cite{K.Hirota98} studied the magnetic structure LSMO327 for
$x=0.40, 0.45$ and $0.48$ by neutron diffraction using single crystals.  They  have
found that the magnetic structure is not simple ferromagnetic (FM) but  planar canted
antiferromagnetic (AFM) and that the canting angle increases from 6.3$^{\circ}$
($x=0.40)$ to 180$^{\circ}$ ($x=0.48)$ while $T_{C}$ decreases with increasing $x$. 
They also pointed out the existence of A-type AFM  ordering in the intermediate
temperature range above $T_{C}$.  Fujioka {\it et al.}\cite{H.Fujioka98} measured the
spin wave dispersion of a large single-grain single crystal of LSMO327 with $x=0.40$
and found that the intra-plane coupling between a bilayer is about 30~\% of the
in-plane coupling ($D \sim 151$~meV\AA) though the Mn-O bond lengths are similar, which
they ascribed to the dominance of $d_{x^{2}-y^{2}}$ orbital.  

The importance of the $e_{g}$ orbital degree of freedom has been pointed out
theoretically as to the change of the magnetic structure in
manganites.\cite{Solo96,S.Ishihara97i}  Note that the $e_{g}$
oribital state is governed by the Jahn-Teller (JT) type distortion of the MnO$_{6}$
octahedron.  Since  LSMO327 ($0.30 \le x \le 0.50$) has neither buckling nor structural
phase transition unlike the three dimensional (3d) Mn Perovskite such as
La$_{1-x}$Sr$_{x}$MnO$_{3}$, we can study more directly how the change of the $e_{g}$
orbital state affects the magnetic structure.  In this Communication,  we have
established by neutron diffraction experiment close interrelation between the  orbital
state and the magnetic structure in the layered manganites LSMO327.


The prescribed amount of dried La$_2$O$_3$, SrCO$_3$, and Mn$_3$O$_4$ are thoroughly
mixed and calcined in the air at 1200--1450$^{\circ}$C for total 4 days with frequent
grinding.  Sample rods were melt-grown in a floating-zone optical image furnace, then
powderized again.  No detectable impurities were found in these samples at room
temperature (R.T.) by x-ray diffraction.  As for
$x=0.30$, inductively coupled plasma (ICP) analysis shows that the ratio of  La, Sr and
Mn is $28.6:32.2:39.2$, which is in good agreement with the ideal ratio
$28.0:32.0:40.0$.  We also checked an $x=0.40$ single crystal
with electron probe microanalysis (EPMA), which indicates no particular spatial
inhomogeneity within the instrumental error.  These analyses indicate that our samples
are sufficiently stoichiometric and homogeneous.  The crystal structures were  studied
as a function of $x$ as well as temperature using the powder neutron  diffractometer
HERMES,  which is equipped with multi-detectors with the Ge $ (3\ 3\ 1)$ monochromator 
($\lambda=1.819$~\AA), located in the JRR-3M reactor in JAERI.\cite{K.Ohoyama}  We
have taken powder diffraction patterns at R.T.  and low temperature ($\sim 10$~K).  As
described in Ref.~\onlinecite{Hirota99}, the peak profiles as well as the magnetic
order parameters indicate that none of our samples show biphasic properties as
previously reported for $x=0.50$ by Battle {\it et al.}\cite{Battle97}


The powder diffraction patterns were analyzed with the Rietveld
method for $x=0.30$, 0.35, 0.40, 0.45, 0.48, and 0.50 at R.T. and at
10~K.  We used melt-grown samples except $x=0.35$.  We confirmed that $T_{C}$ and the
results of Rietveld analyses  of the calcined $x=0.40$ sample  are consistent with
those of the  melted $x=0.40$ sample.   Therefore using the calcined $x=0.35$ sample
should have no fundamental problems in terms of a systematic study of a relation 
between the structural and magnetic structures.  We found that 
LSMO327 belongs to the space group $I4/mmm$ ($z=2$) over the
whole hole concentration range  ($0.30 \leq x
\leq0.50$).  The results of Rietveld analyses are summarized in Tables I and II. 

Let us first mention the magnetic phase diagram of LSMO327 shown in Figs.~1(a) and (b). 
We have found that the low-temperature magnetic structure for $0.32 \le x \le 0.48$ is
essentially classified into a single magnetic phase which consists of FM (FM-I: planar
FM) and AFM (AFM-I: intra-bilayer AFM and inter-layer FM) components as previously
reported for $x=0.40, 0.45, 0.48$ single crystals.\cite{K.Hirota98}  Note that the
AFM-I component exists only above $x=0.39$.  The intermediate AFM-I phase was observed
for $0.39 \le x \le 0.48$.  For $x=0.30$, a different planar AFM structure appears in
the intermediate temperature range, which we denote AFM-II (intra-bilayer AFM and
inter-layer FM).  At low temperature, a new FM phase with the magnetic easy-axis
parallel to $c$ appears.  We define this phase as FM-II.  At the other end,
$x=0.50$, the magnetic structure shows fairly complicated temperature dependence
because of charge ordering.\cite{Li98,M.Kubota98ii}  Neutron-diffraction study on an
$x=0.50$ single crystal revealed that the CE-type charge and A-type AFM ordering occurs
below 210~K and that the CE-type AFM coexists the A-type AFM below 145~K, though the
CE-type AFM and charge ordering become unstable at low temperature.  Present powder
diffraction study is consistent with Ref.~\onlinecite{M.Kubota98ii}. 
Detailed powder patterns are published elsewhere.\cite{M.Kubota98} 

Figure~1(c) show the $x$ dependence of magnetic moments.  Only the FM
component exists for $x=0.30-0.38$ at 10~K.  However
the FM component starts decreasing on the appearance of the AFM component at
$x=0.39$.  This makes a sharp contrast with the case of the 3d Mn Perovskite
La$_{1-x}$Sr$_{x}$MnO$_{3}$, in which FM structure exists over $0.17<x<0.50$.  As the
AFM component increases as $x$ increases, the FM component diminishes and finally
disappears at $x= 0.48$.  These results
are consistent with neutron diffraction studies using single
crystals.\cite{K.Hirota98}  Figure~1(d) describes the JT distortion
$\Delta_{JT}$ defined as the ratio of the averaged apical\cite{apical} and the
equatorial Mn-O bond lengths at 10~K and R.T.  At R.T., the lattice constant $c$
greatly decreases with increasing $x$, while the  lattice constant $a$ stays almost
constant (See Table~1); $\frac{c(x=0.50)-c(x=0.30)}{c(x=0.50)}$ is $-1.46$~\%, while
$\frac{a(x=0.50)-a(x=0.30)}{a(x=0.50)}$ is only 0.38~\%.  At 10 K, the
former value is $-1.61$~\% and the latter is $0.45$~\% (See Table~II).  Thus the lattice
shrinks along $c$ as $x$ increases.  On the other hand, the out-of-plane Mn-O-Mn angle
$\Theta_{a}$ is found almost 180$^{\circ}$ independent of $x$, indicating no buckling. 
The angle along $c$, $\Theta_{c}$, is 180$^{\circ}$ because of the $I4/mmm (z=2)$
tetragonal symmetry. As expected from the $x$ dependence of the lattice constants $a$
and $c$, the bond length Mn-O$_{equatorial}$ in MnO$_6$ octahedra stays almost
constant and the bond length Mn-O$_{apical}$ largely decreases
with increasing $x$.  Thus $\Delta_{JT}$ changes from 1.042(5) ($x=0.30$) to 1.010(6)
($x=0.50$) at R.T.  At 10~K, the former value is 1.034(6) and the latter is 1.010(5). 
In the high dope region ($0.45 \leq x \leq 0.50$), the JT distortion is so small that
MnO$_6$ octahedron is almost regular ($\Delta_{JT}=1$).


In manganites, FM double-exchange (DE) interaction, mediated by carrier hopping, and AFM
superexchange (SE) interaction compete with
each other.\cite{P.W.Anderson55,P.G.deGennes60}  The change of the $e_{g}$ orbital
state through the structural parameters should modify the anisotropic transfer
integral, and affect the above competition.  The $d_{x^2-y^2}$ orbital, which are 
theoretically more stable than $d_{3z^2-r^2}$ orbitals in LSMO327,\cite{S.Ishihara97ii}
reflecting the two-dimensional character, facilitates the FM DE
interactions within the plane, while interactions along $c$ are dominated by the AFM
SE coupling between $t_{2g}$ spins.  Consequently the layered A-type AFM structure
prevails over 3d FM structure, making a sharp contrast with cubic Perovskites like
La$_{1-x}$Sr$_{x}$MnO$_{3}$.  As shown in Figs.~1(c) and (d), we have found a clear
correlation between the decrease of $\Delta_{JT}$ and the crossover from FM to AFM
component for $0.32 \leq x \leq 0.48$.  Here, the contraction of $\Delta_{JT}$ with
increasing $x$ stabilizes the energy level of the $d_{x^2-y^2}$ orbital.  This
observation indicates that the increase of $d_{x^2-y^2}$ orbital polarization
enhances the A-type AFM component. 

We should also point out a compositional phase boundary between $x=0.30$
and $0.32$, which is clearly indicated by an abrupt change in the direction 
of the easy axis though the structural parameters change
continuously.  Note that different magnetic structures have been reported for $x=0.30$. 
Perring {\it et al.} \cite{Perring98} proposed a low-temperature AFM structure, which is
intra-bilayer FM and inter-bilayer AFM coupling with the easy axis parallel to $c$
(AFM-III in Fig.~1(b)).  Argyriou {\it et al.}\cite{Argyriou99} reported that their
$x=0.30$ phase is biphasic and that the majority phase (hole poor) is a canted AFM
structure and the minority phase (hole rich) is a tilted FM structure with
inter-bilayer FM coupling.  The tilt angle approaches 0$^{\circ}$, i.e., parallel to
$c$, as temperature decreases for the both phases.  Thus the majority phase
becomes AFM-III as suggested by Perring {\it et al.}\ and the minority phases becomes
FM-II consistent with the present study.  These results indicate that there exist at
least two low-temperature magnetic phases (FM-II and AFM-III) around $x=0.30$.  As
Moritomo {\it et al.}\cite{Moritomo98ii} concluded in their study of
chemical pressure effects on the magnetotransport properties 
of Nd-doped LSMO327, the complicated  magnetic phases around $x=0.30$ as above result
from a subtle balance between inter-bilayer FM and AFM coupling, thus is sensitive to
the hole concentration controlling the magentic interactions through structural
parameters.

In summary, we have determined the crystal and magnetic structure of layered
manganites  La$_{2-2x}$Sr$_{1+2x}$Mn$_{2}$O$_{7}$, and established the close relation
between the JT distortion and the magnetic structure. Continuous variation from FM to
canted AFM structure with increasing $x$ ($0.32 - 0.48$) is interpreted in terms
of increase of  the $d_{x^2-y^2}$ orbital polarization associated with decrease of the
JT distortion.

\begin{figure}
\begin{center}
\BoxedEPSF{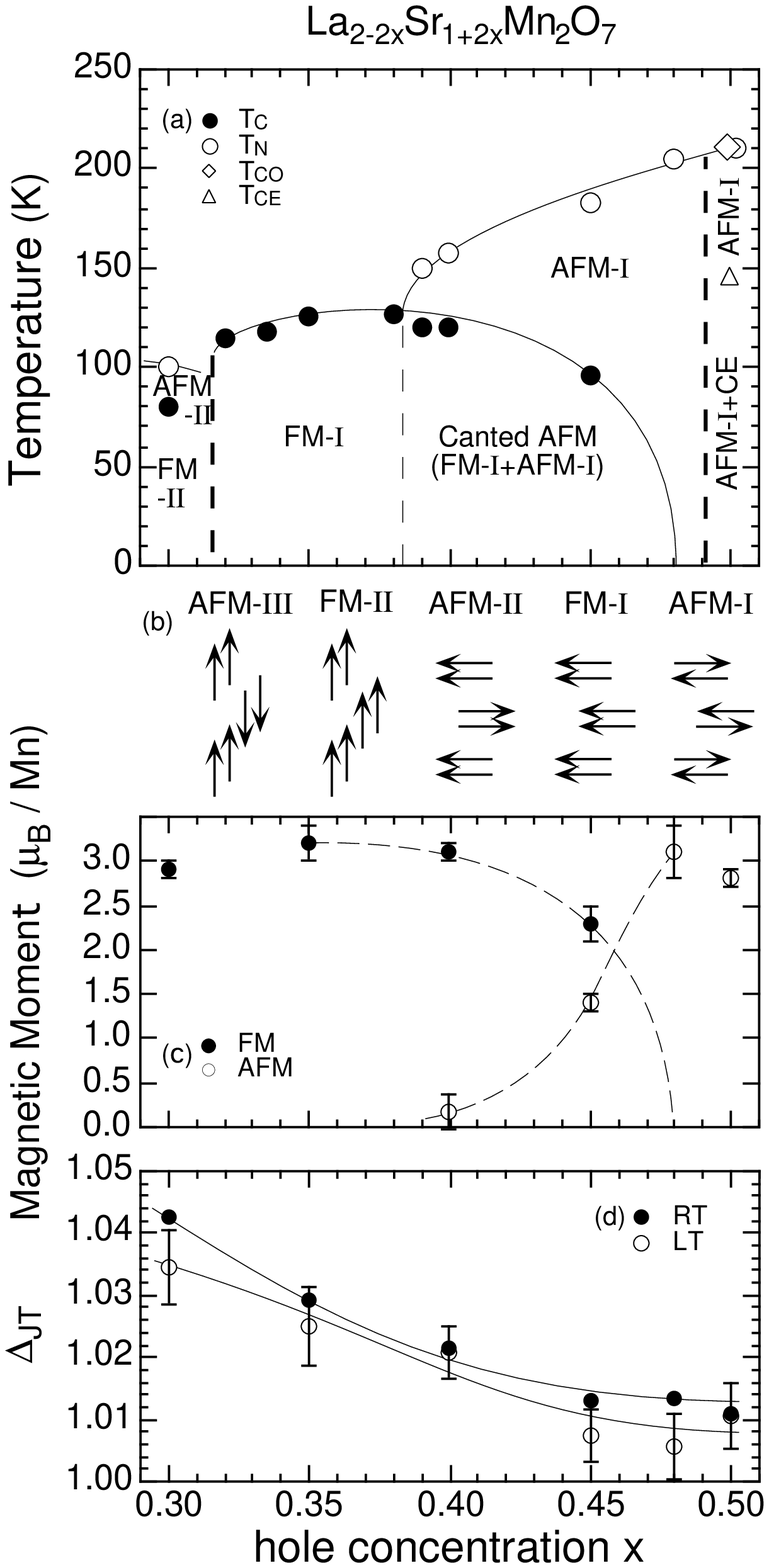 scaled 600}
\vspace{0.5cm}
\end{center}
\caption{(a) Magnetic phase diagram of La$_{2-x}$Sr$_{1+2x}$Mn$_{2}$O$_{7}$ 
($0.30 \leq x \leq 0.50$). (b) Several different magnetic structures appearing in
the phase diagram are schematically drawn.  AFM-III is reported in
Ref.~\protect\onlinecite{Perring98}.  (c) Hole concentration dependence of FM and AFM
mangetic moments at 10~K.  (d) Hole concentration dependence of JT distortion defined
as the ratio of the averaged apical and the equatorial Mn-O bond lengths at room
temperature and 10~K.}
\label{T-xphase}
\end{figure}

\twocolumn[\hsize\textwidth\columnwidth\hsize\csname @twocolumnfalse\endcsname

\begin{table}
\caption{Structural parameters for La$_{2-2x}$Sr$_{1+2x}$Mn$_2$O$_7$ at room
temperature.  In the space group {\it I}4/{\it mmm}, the Mn is placed at (0,
 0, z), La,Sr(1) at (0, 0, 0.5),  La,Sr(2) at (0, 0, z), O(1) at (0, 0, 0), O(2)
 at (0, 0, z), O(3) at (0, 0.5, z), respectively. 
$\Theta_{a}$ and  $\Delta_{JT}$ denote the out-of-plane angle between Mn and 
O(3) and the Jahn-Teller distortion, respectively.  $R_{wp}$ is a weighted pattern
R-factor and  $R_{e}$ is an expected R-factor.  Non-melted samples were used for
$x=0.35^*$ and 0.40$^*$.}\label{mag-ana}
\begin{center}
\begin{tabular}{lllllllll}           &      & $x=0.30$       & $x=0.35^*$   
& $x=0.40$   & $x=0.40^*$ & $x=0.45$ & $x=0.48$ & $x=0.50$ \\ \hline &$a$ (\AA)
&3.8600(1) &3.8652(3) & 3.8711(1)& 3.8703(2)  & 3.8729(2) & 3.8765(2) &
3.8748(3)  \\  &$c$ (\AA) & 20.324(1)& 20.177(2) & 20.126(1)& 20.110(1)  &
20.056(1) & 20.049(1) & 20.032(1) \\  &Mn        & 0.097(1) &0.097(1)  
&0.097(1) & 0.097(1)& 0.097(1)& 0.097(1)  &0.098(1) \\  &La,Sr(2) &  0.3172(7)&
0.3174(8)& 0.3174(6)&0.3170(7)  &0.3171(6)&0.3168(6)&0.3161(7)  \\  &O(2) &
0.1980(9)& 0.197(1)& 0.1965(8)&0.196(1) & 0.1956(8)& 0.1960(9)& 0.195(1) \\ 
 &
O(3) &0.0958(6)&0.0946(7)&0.0957(5)&0.0955(6) &0.0949(5)&0.0948(5)&0.0944(6)
 \\ 
&$\Theta_{a}$ (deg)&177.7(4)&176.7(5)&178.0(4)&178.3(4) 
&177.1(3)&176.7(4)&175.3(4) \\ 
&$\Delta_{JT}$&1.042(5)&1.029(8)&1.021(4)&1.019(5)&1.013(4)&1.013(5)&1.010(6)\\
&$R_{wp}$ &10.23&11.69&9.41&12.82&9.28&10.14&11.13 \\
&$R_{e}$ &3.92&4.50&4.85&6.02&4.62&4.94&4.51 \\
\end{tabular}
\end{center}
\end{table}

\begin{table}
\caption{Structural parameters for La$_{2-2x}$Sr$_{1+2x}$Mn$_2$O$_7$ at 10~K. 
$\mu_F$ and $\mu_{AF}$ indicate ferromagnetic moment and antiferromagnetic
moment  per Mn site.  Non-melted samples were used for $x=0.35^*$.}
\label{mag-ana}
\begin{center}
\begin{tabular}{llllllll}           &      & $x=0.30$ &  $x=0.35^*$ & $x=0.
40$ &
$x=0.45$ & $x=0.48$ & $x=0.50$ \\ \hline &$a$ (\AA) &3.8556(4)&3.8587(2) &
3.8645(1)  & 3.8713(2) & 3.8782(1) & 3.8731(3)  \\  &$c$ (\AA) &
20.239(2)&20.096(2) & 20.065(1)  & 19.963(1) & 19.952(1) & 19.918(2) \\ 
&Mn        & 0.097(1)&0.096(1)  &0.096(1)& 0.097(1)& 0.097(1)  &0.097(1) \\ 
&La,Sr(2)  &  0.3182(8)&0.3173(7) & 0.3178(6)&0.3165(6)&0.3174(8)&0.3174(7) 
 \\ 
&O(2)      & 0.197(1)& 0.196(1) & 0.1966(8)& 0.1954(8)& 0.195(1)& 0.1965(9) 
\\ 
& O(3)     &0.0955(7)& 0.0949(7) &0.0957(5)&0.0951(5)&0.0949(7)&0.0949(6) \\
 
&$\Theta_{a}$ (deg)&177.9(5)&177.9(5)  &178.9(3)&177.7(3)&178.0(5)&177.0(4)
 \\
& $\mu_F$ &2.9(2)&3.2(2)&3.1(1)&2.3(2)&---&--- \\ & $\mu_{AF}$ &---
&---&0.17(2)&1.4(1)&3.1(3)&2.8(1) \\  
&$\Delta_{JT}$&1.034(6)&1.025(6)&1.021(4)&1.008(4)&1.006(5)&1.010(5)\\
&$R_{wp}$ &13.07&12.21&11.94&12.21&13.35&13.16 \\
&$R_{e}$ &5.12&4.80&5.26&4.81&5.38&5.01 \\
\end{tabular}
\end{center}
\end{table}

]

\end{document}